\documentclass[sigconf,nonacm]{acmart}

\usepackage{graphicx}
\usepackage{xcolor}
\usepackage{xspace}
\usepackage[acronym,nohypertypes={acronym}]{glossaries}
\usepackage{hyperref}

\newacronym{iaas}{\textsc{I}aa\textsc{S}}{\textit{Infrastructure-as-a-Service}}
\newacronym{vm}{VM}{\textit{Virtual Machine}}
\newacronym{qos}{QoS}{\textit{Quality of Service}}
\newacronym{nic}{NIC}{\textit{Network Interface Card}}
\newacronym{cdf}{CDF}{\textit{Cumulative Distribution Function}}

\begin{document}

\title{Beyond Throttling: The Case for Dynamic NIC Oversubscription in IaaS Clouds}

\author{Pierre Jacquet}
\orcid{0009-0002-7988-8550}
\affiliation{%
  \institution{École de Technologie Supérieure, Université du Québec}
  \city{Montréal}
  \country{Canada}}
\email{pierre.jacquet@etsmtl.ca}

\author{Camille Coti}
\orcid{0000-0002-1224-7786}
\affiliation{%
  \institution{École de Technologie Supérieure, Université du Québec}
  \city{Montréal}
  \country{Canada}}
\email{camille.coti@etsmtl.ca}

\author{Marcos Dias De Assunção}
\orcid{0000-0002-4218-0260}
\affiliation{%
  \institution{École de Technologie Supérieure, Université du Québec}
  \city{Montréal}
  \country{Canada}}
\email{marcos.dias-de-assuncao@etsmtl.ca}

\renewcommand{\shortauthors}{Jacquet et al.}

\begin{abstract}
Cloud platforms routinely oversubscribe resources to exploit the gap between allocated and actual usage.
Network bandwidth is a notable exception: because network activity is bursty and hard to predict, providers instead expose instances with a variable \gls{nic} bandwidth governed by credit-based throttling, which is opaque to users and can limit throughput even when the host has spare capacity.
In this paper, we argue that \gls{nic} bandwidth oversubscription should be handled by the scheduler rather than by throttling.
We characterize network usage in a production \gls{iaas} infrastructure operated by OVHcloud, covering more than 75,000 \glspl{vm}.
We find that \gls{nic} capacity is largely unused, that individual \gls{vm} peaks cannot be inferred from \gls{vm} attributes, and that host-level peaks are short-lived, rare, and highly heterogeneous across hosts, such that no single static oversubscription ratio is safe.
Building on these observations, we outline a dynamic \gls{nic} filter: a placement-time admission check that estimates each host's future peak demand from observations pooled across similar hosts, allowing instances to keep a constant bandwidth allocation while the physical capacity is safely shared.
\end{abstract}


\maketitle
\glsresetall

\section{Introduction}\label{sec:intro}

\Glspl{vm} rarely use all of their allocated resources.
Cloud providers exploit this gap through resource oversubscription\footnote{Also referred to as overcommitment.}, serving more virtual resources than the infrastructure can simultaneously provide, which lowers both infrastructure costs and the environmental footprint of cloud computing~\cite{10.1145/3132747.3132772,266915,258850,verma2015large}.
Because each resource type has distinct usage patterns, oversubscription requires mechanisms tailored to each of them.
CPU and memory oversubscription have been studied extensively~\cite{10.1145/3447786.3456225,DBLP:journals/tsusc/JacquetLR24,10.1145/3447786.3456259,10.1145/2007477.1952709,jennings2015resource}, and storage has more recently received dedicated attention~\cite{10.1145/3669940.3707226,sc26disk}.
Network bandwidth, another fundamental resource, has followed a markedly different path.

Traffic leaving or entering an instance crosses two kinds of devices: the host \gls{nic} and the switches forwarding packets toward their destination.
Switches are inherently oversubscribed, and a large body of work studies how to allocate and guarantee bandwidth across them.
The host \gls{nic}, in contrast, is rarely managed as an oversubscribed resource.
Instead, most public clouds rely on reactive, credit-based throttling: an instance receives a peak bandwidth and a lower baseline, and its achievable throughput moves between the two depending on accumulated credits.
This design accommodates the bursty nature of network traffic without provisioning peak capacity for every instance, but it has clear drawbacks: its behavior is opaque to users, network performance varies over time, and an instance may be throttled even when the host \gls{nic} has spare capacity.

We ask whether \gls{iaas} platforms can instead offer instances a \emph{constant and predictable} \gls{nic} bandwidth while still oversubscribing the physical capacity.
Our position is that the uncertainty of network demand should be absorbed by the scheduler when placing instances, rather than by throttling instances once placed.
This paper makes two contributions:
\begin{itemize}
\item We characterize network activity in production \gls{iaas} clusters operated by OVHcloud at the \gls{vm}, host, and cluster levels. We show that \gls{nic} capacity is largely unused, but that network peaks are short-lived, poorly correlated with \gls{vm} attributes, and highly heterogeneous across hosts.
\item We derive from these observations the principles of a \emph{dynamic \gls{nic} filter}, a placement-time admission check that replaces static oversubscription ratios and credit-based throttling by host-specific peak estimates learned from pooled observations.
\end{itemize}

Section~\ref{sec:rw} reviews related work, Section~\ref{sec:motiv} presents our characterization, Section~\ref{sec:filter} outlines the dynamic filter, and Section~\ref{sec:conclusion} concludes.

\section{Related Work}\label{sec:rw}

\paragraph{Cloud networking.}
Production network virtualization stacks such as Andromeda~\cite{dalton2018andromeda}, VFP~\cite{201555}, and Achelous~\cite{wei2023achelous} map virtual networks onto physical infrastructure, and have been optimized for latency~\cite{10.1145/3651890.3672224,10.1145/3689031.3717479} and scale~\cite{295587,dalton2018andromeda}.
We do not modify this stack; we study how observed network usage can inform how much \gls{nic} bandwidth is placed on each host.

\paragraph{Switch-level provisioning.}
Since the aggregate bandwidth below a switch typically exceeds its uplink capacity, bandwidth allocation must account for the network hierarchy.
SecondNet~\cite{10.1145/1921168.1921188} and Oktopus~\cite{10.1145/2043164.2018465} reserve bandwidth from explicit tenant specifications, MAPLE~\cite{MAPL} derives effective bandwidth from observed traffic, and TIVC~\cite{TIVC} accounts for its temporal variations.
These approaches address contention along network paths, whereas we consider a different allocation boundary: the aggregate instance bandwidth that can be placed behind a single host \gls{nic}.

\begin{figure}[t]
\centering
\includegraphics[width=0.85\linewidth]{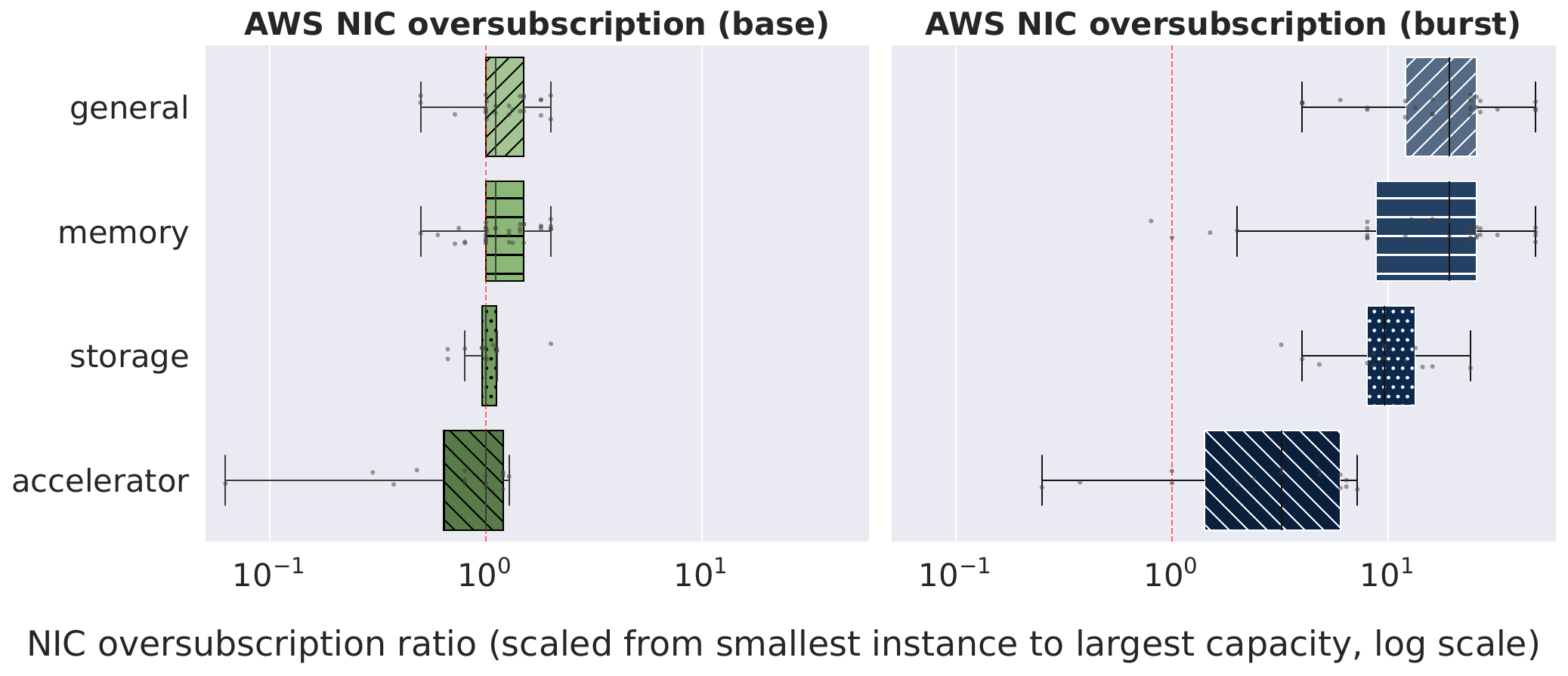}
\caption{AWS NIC bandwidth oversubscription ratio (base and burst) across instance families (log scale). The red dashed line marks a ratio of 1 (no oversubscription).}
\Description{AWS NIC bandwidth oversubscription per flavor family. Burst values are highly oversubscribed while baseline values correspond to no oversubscription.}
\label{fig:rw:aws}
\end{figure}

\paragraph{NIC-level provisioning.}
Providers widely advertise instance bandwidth above what can be simultaneously guaranteed.
AWS\footnote{Selected for its detailed instance documentation and market share: \url{https://docs.aws.amazon.com/ec2/latest/instancetypes/ec2-instance-type-specifications.html}}, for instance, specifies a baseline and a higher burst bandwidth for its instance types.
Comparing the network capacity of the largest instance of a family, which corresponds to a full host~\cite{sizingaws}, with the aggregate burst bandwidth of the equivalent number of smallest instances, we estimate a median \gls{nic} oversubscription of $19.2{:}1$ for general-purpose families, against only $1.1{:}1$ for baseline bandwidth (Figure~\ref{fig:rw:aws}).
Systems such as SENIC~\cite{radhakrishnan2014senic} and Carousel~\cite{saeed2017carousel} manage the resulting contention at the host, and several studies report the performance variability users experience in public cloud networks~\cite{10.1145/3570609,10196385,9718592}.
These approaches, like credit-based throttling, react to demand after instances are placed.
By contrast, while CPU~\cite{10.1145/3132747.3132772,10.1145/3669940.3707226,DBLP:journals/tsusc/JacquetLR24}, memory~\cite{fuerst2022memory}, and storage~\cite{sc26disk} oversubscription have been framed as dynamic placement problems, \gls{nic} bandwidth has received little attention as such.

\section{Production Network Usage}\label{sec:motiv}

We characterize network usage in a commercial \gls{iaas} cluster comprising more than 75,000 \glspl{vm} over 7,500 host-days.
Earlier datacenter traffic studies~\cite{10.1145/1879141.1879175,10.1145/1672308.1672325,alizadeh2010data} are more than 15 years old, predate substantial hardware and \gls{vm} evolutions~\cite{cloudevolutionccgrid26}, and focus on aggregate traffic rather than per-instance usage.
OVHcloud Public Cloud runs OpenStack with QEMU/KVM hypervisors on a two-tier network.
Two properties make it well suited to observe network demand: it applies no burst-credit throttling to instances, and its EU regions neither charge for traffic nor impose quotas, limiting incentives to suppress network usage.

\subsection{VM View}

\begin{figure}[t]
\centering
\includegraphics[width=0.95\linewidth]{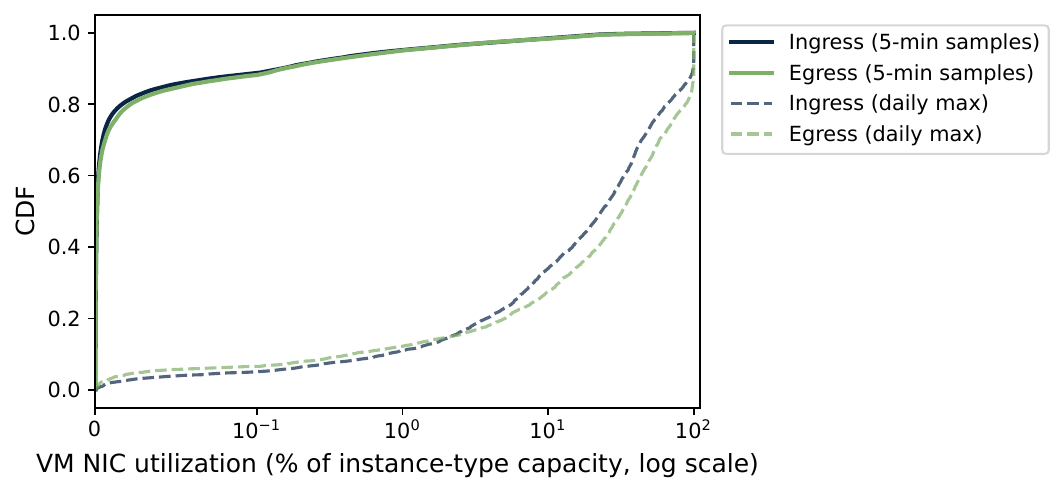}
\caption{CDF of VM ingress and egress utilization relative to provisioned NIC capacity: 5-minute averages (solid) and daily maxima (dashed).}
\Description{CDF of VM ingress and egress bandwidth utilization relative to the provisioned NIC capacity.}
\label{fig:motiv:vm-cdf}
\end{figure}

Figure~\ref{fig:motiv:vm-cdf} shows \gls{vm} bandwidth utilization relative to the provisioned capacity.
Average utilization is extremely low: about 80\% of 5-minute samples remain below 0.1\% of the provisioned bandwidth.
Daily peaks tell a different story: 80\% of \glspl{vm} exceed 10\% of their allocation at least once a day.
Bandwidth therefore cannot simply be reduced to a small static allocation, but aggregate demand remains far below provisioned capacity, leaving room for statistical multiplexing.

Can a scheduler anticipate these peaks from the information it has at placement time?
The Spearman correlation between \gls{vm} attributes (vCPU count, allocated bandwidth) and daily peak ingress or egress traffic is weak, the highest coefficient being 0.31 (vCPUs vs.\ egress), and Pearson correlation yields similar results.
\Gls{vm} attributes thus provide little information about future network peaks.

\subsection{Host View}

\begin{figure}[t]
\centering
\includegraphics[width=0.95\linewidth]{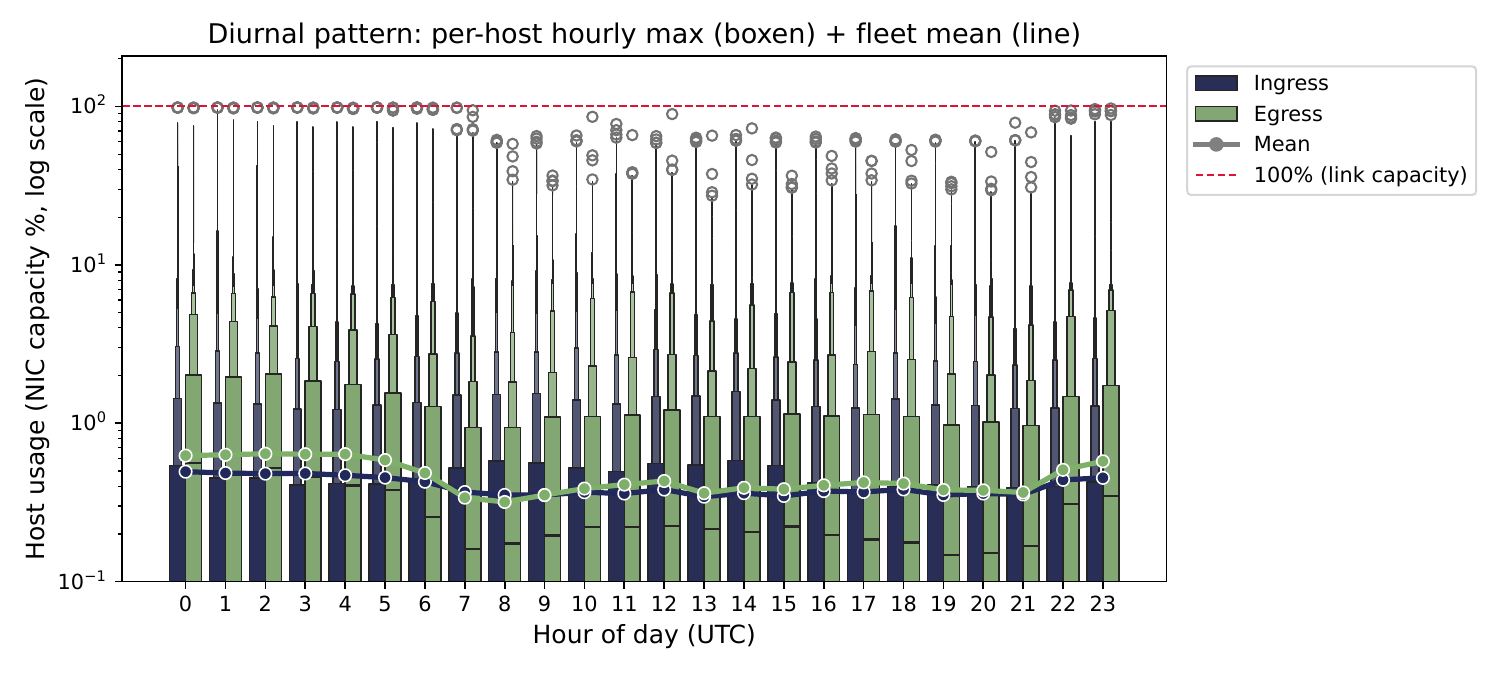}
\caption{Hourly maximum NIC utilization across hosts: mean (line) and dispersion (boxplots), log scale.}
\Description{Hourly maximum NIC utilization across hosts over time. NICs have peaks, while remaining mostly idle.}
\label{fig:motiv:cluster-nic-usage}
\end{figure}

At the host level, the scheduler can rely on multiplexing across many \glspl{vm}.
Figure~\ref{fig:motiv:cluster-nic-usage} shows that most hosts operate at low \gls{nic} utilization, with a slight diurnal pattern and substantial dispersion between hosts, including a few high-utilization outliers.
These peaks are short-lived and can be orders of magnitude above typical usage.
Unlike CPU, memory, or storage, whose utilization is sustained and can be characterized from ongoing activity, network capacity requirements are dominated by rare bursts.
Extremes are thus intrinsically sparse: a host yields roughly one meaningful maximum per day, and a long history is not an option since host composition changes and many instances are short-lived~\cite{10.1145/3132747.3132772}.
Estimating the tail of each host independently is therefore impractical.

\subsection{Cluster View}

Pooling observations across hosts yields more samples, but raw utilization is not directly comparable across hosts.
We illustrate this heterogeneity using daily maxima of 5-minute egress observations, considering a \gls{vm} active when it exceeds 10\% of its allocated bandwidth.

\paragraph{Concurrency.}
High ratios of simultaneously active \glspl{vm} are mostly found on hosts with few, typically large \glspl{vm}~\cite{jacquet2023cloudfactory}; as more domains share a host, simultaneous activity becomes less likely, helped by anti-affinity rules spreading instances of a same domain.
Yet hosts span from 1 to more than 140 domains, and significant concurrency still occurs on some densely populated hosts, which may become contention points once the \gls{nic} is oversubscribed.

\paragraph{Implied oversubscription.}
Weighting each instance by its size, we express each host's peak usage as a fraction of its allocated \gls{nic} capacity.
98\% of host peaks remain below half of the allocation, suggesting that a $2\times$ oversubscription would suit most hosts.
However, only 46\% of hosts would support a $20\times$ ratio, close to the one inferred from AWS burst bandwidth.
Substantial oversubscription is thus possible, but peak activity is too heterogeneous to be handled by a single static ratio.

\paragraph{Absolute vs.\ relative peaks.}
Splitting hosts into two groups according to their 95th-percentile egress utilization, the cold group, holding 99\% of hosts, has tightly concentrated absolute peaks (median 10\%), while the hot group is higher and more dispersed.
Normalizing instead each host's observations by the mean $\mu_h$ and variability $\sigma_h$ expected from its current \glspl{vm}, as $z_h(t) = (u_h(t)-\mu_h(t))/\sigma_h(t)$, the cold group shows substantial variation in its extremes despite its concentrated absolute peaks.
Absolute peak utilization therefore does not characterize the risk of adding capacity to a host: the same peak carries different meaning on hosts with different expected demand.

\subsection{Takeaways}
\begin{enumerate}
\item[\textbf{T1}] \gls{nic} capacity is largely unused, leaving substantial room for oversubscription.
\item[\textbf{T2}] Individual \gls{vm} peaks cannot be predicted from \gls{vm} attributes.
\item[\textbf{T3}] Host peaks are rare and short-lived, so each host provides too few extremes to estimate its own tail.
\item[\textbf{T4}] Hosts are heterogeneous: no single oversubscription ratio is safe, and observations are only comparable once expressed relative to each host's expected demand.
\end{enumerate}

\section{Towards a Dynamic NIC Filter}\label{sec:filter}

These observations suggest moving \gls{nic} oversubscription from the data path to the control plane.
Cloud orchestrators such as OpenStack, Borg~\cite{verma2015large}, or Protean~\cite{258979} place instances through a filter/weigher pipeline~\cite{openstacksched}: filters discard hosts that violate hard constraints, and weighers rank the remaining candidates.
Today, \gls{nic} bandwidth is either ignored by this pipeline or checked against a static oversubscription ratio, and contention is left to throttling.
We instead propose to treat \gls{nic} bandwidth as a \emph{dynamic} scheduling constraint.

\paragraph{Principle.}
Each instance is advertised a constant bandwidth, with no baseline or credits.
When placing a new \gls{vm}, a dynamic \gls{nic} filter estimates the peak demand the host would reach with the candidate added, and admits the placement only if this estimate remains below the physical \gls{nic} capacity.
The effective oversubscription ratio is thus no longer a fleet-wide constant: it emerges per host from observed activity, high on hosts whose \glspl{vm} rarely burst together and lower on those exhibiting concurrent peaks (T1, T4).

\paragraph{Design requirements.}
Our characterization translates into the following requirements for the peak estimate:
\begin{itemize}
\item \emph{Composition-based prior.} Since a newly placed \gls{vm} has no history and its attributes are poor predictors (T2), the expected demand of a host should be derived from the usage profiles observed for the offerings of its hosted \glspl{vm}, updated as the composition changes.
\item \emph{Tail-oriented.} Capacity is driven by rare bursts rather than averages (T3), so the filter must target an upper quantile of host demand rather than its mean or a fixed percentile of past usage.
\item \emph{Pooled and host-relative.} Because a single host yields too few extremes (T3), observations should be pooled across hosts with similar usage regimes, after being normalized by each host's expected demand so that they become comparable (T4).
\item \emph{Saturation-aware.} When a \gls{nic} saturates, measured usage is capped at capacity and understates true demand; such observations must be treated as censored rather than as regular peaks.
\item \emph{Online.} Estimates must be refreshed continuously from monitoring data, without requiring changes to the guest or to the advertised allocation.
\end{itemize}

\paragraph{Complementary placement.}
The filter decides \emph{whether} a host can accept an instance; weighers can further decide \emph{where} it is best placed.
Since concurrent activity concentrates on hosts with few large \glspl{vm} (Section~\ref{sec:motiv}), weighers can favor placements that diversify independent \gls{nic} allocations, \emph{e.g.,} spreading large \glspl{vm} and steering them away from hosts with high observed usage.

\paragraph{Trade-offs.}
A dynamic filter turns oversubscription into an explicit provider trade-off between packing density and \gls{qos}: a conservative estimate rejects placements that a static ratio would accept, but avoids exposing users to unpredictable bandwidth loss.
We favor proactive placement over live migration to resolve contention, since migrations themselves consume bandwidth on an already oversubscribed network.
The design and evaluation of a concrete filter following these principles, using production network traces, is the subject of ongoing work.

\section{Conclusion}\label{sec:conclusion}

Unlike CPU and memory, \gls{nic} bandwidth in public clouds is typically shared through credit-based throttling rather than scheduler-driven oversubscription.
Our characterization of a production OVHcloud \gls{iaas} infrastructure shows that \gls{nic} capacity is largely unused, but that network peaks are bursty, poorly predictable from \gls{vm} attributes, and heterogeneous across hosts.
We argued that these properties call for a dynamic \gls{nic} filter that estimates host-specific peak demand from pooled, host-relative observations, allowing providers to offer stable bandwidth to instances while safely sharing physical \gls{nic} capacity.

\bibliographystyle{ieeetr}
\bibliography{biblio}

@inproceedings{radhakrishnan2014senic,
  title={$\{$SENIC$\}$: Scalable $\{$NIC$\}$ for $\{$End-Host$\}$ rate limiting},
  author={Radhakrishnan, Sivasankar and Geng, Yilong and Jeyakumar, Vimalkumar and Kabbani, Abdul and Porter, George and Vahdat, Amin},
  booktitle={11th USENIX Symposium on Networked Systems Design and Implementation (NSDI 14)},
  pages={475--488},
  year={2014}
}

@inproceedings{10.1145/1921168.1921188,
author = {Guo, Chuanxiong and Lu, Guohan and Wang, Helen J. and Yang, Shuang and Kong, Chao and Sun, Peng and Wu, Wenfei and Zhang, Yongguang},
title = {SecondNet: a data center network virtualization architecture with bandwidth guarantees},
year = {2010},
isbn = {9781450304481},
publisher = {Association for Computing Machinery},
address = {New York, NY, USA},
url = {https://doi.org/10.1145/1921168.1921188},
doi = {10.1145/1921168.1921188},
booktitle = {Proceedings of the 6th International COnference},
articleno = {15},
numpages = {12},
location = {Philadelphia, Pennsylvania},
series = {Co-NEXT '10}
}

@inproceedings{10.1145/1879141.1879175,
author = {Benson, Theophilus and Akella, Aditya and Maltz, David A.},
title = {Network traffic characteristics of data centers in the wild},
year = {2010},
isbn = {9781450304832},
publisher = {Association for Computing Machinery},
address = {New York, NY, USA},
url = {https://doi.org/10.1145/1879141.1879175},
doi = {10.1145/1879141.1879175},
booktitle = {Proceedings of the 10th ACM SIGCOMM Conference on Internet Measurement},
pages = {267–280},
numpages = {14},
location = {Melbourne, Australia},
series = {IMC '10}
}

@article{10.1145/1672308.1672325,
author = {Benson, Theophilus and Anand, Ashok and Akella, Aditya and Zhang, Ming},
title = {Understanding data center traffic characteristics},
year = {2010},
issue_date = {January 2010},
publisher = {Association for Computing Machinery},
address = {New York, NY, USA},
volume = {40},
number = {1},
issn = {0146-4833},
url = {https://doi.org/10.1145/1672308.1672325},
doi = {10.1145/1672308.1672325},
journal = {SIGCOMM Comput. Commun. Rev.},
month = jan,
pages = {92–99},
numpages = {8}
}

@ARTICLE{9718592,
  author={Iqbal, Hassan and Singh, Anand and Shahzad, Muhammad},
  journal={IEEE/ACM Transactions on Networking}, 
  title={Characterizing the Availability and Latency in AWS Network From the Perspective of Tenants}, 
  year={2022},
  volume={30},
  number={4},
  pages={1554-1568},
  doi={10.1109/TNET.2022.3148701}}

@article{10.1145/3570609,
author = {De Sensi, Daniele and De Matteis, Tiziano and Taranov, Konstantin and Di Girolamo, Salvatore and Rahn, Tobias and Hoefler, Torsten},
title = {Noise in the Clouds: Influence of Network Performance Variability on Application Scalability},
year = {2022},
issue_date = {December 2022},
publisher = {Association for Computing Machinery},
address = {New York, NY, USA},
volume = {6},
number = {3},
url = {https://doi.org/10.1145/3570609},
doi = {10.1145/3570609},
journal = {Proc. ACM Meas. Anal. Comput. Syst.},
month = dec,
articleno = {49},
numpages = {27}
}

@ARTICLE{10196385,
  author={Yildirim, Esma and Akon, Arafat},
  journal={IEEE Access}, 
  title={Predicting Short-Term Variations in End-to-End Cloud Data Transfer Throughput Using Neural Networks}, 
  year={2023},
  volume={11},
  number={},
  pages={78656-78670},
  doi={10.1109/ACCESS.2023.3299311}}

@inproceedings{cloudevolutionccgrid26,
  TITLE = {{Larger Cloud Servers, Fewer Hosts? On the Evolution of VM Sizes in IaaS Platforms}},
  AUTHOR = {Jacquet, Pierre and Camille Coti and Marcos Dias de Assunção},
  BOOKTITLE = {IEEE International Symposium on Cluster, Cloud and Internet Computing (CCGrid)},
  PUBLISHER = {IEEE},
  year={2026},
  pages={287-296},
  DOI = {10.1109/CCGrid68966.2026.00037},
}

@inproceedings{alizadeh2010data,
  title={Data center tcp (dctcp)},
  author={Alizadeh, Mohammad and Greenberg, Albert and Maltz, David A and Padhye, Jitendra and Patel, Parveen and Prabhakar, Balaji and Sengupta, Sudipta and Sridharan, Murari},
  booktitle={Proceedings of the ACM SIGCOMM 2010 Conference},
  pages={63--74},
  year={2010}
}

@inproceedings {201555,
author = {Daniel Firestone},
title = {{VFP}: A Virtual Switch Platform for Host {SDN} in the Public Cloud},
booktitle = {14th USENIX Symposium on Networked Systems Design and Implementation (NSDI 17)},
year = {2017},
isbn = {978-1-931971-37-9},
address = {Boston, MA},
pages = {315--328},
url = {https://www.usenix.org/conference/nsdi17/technical-sessions/presentation/firestone},
publisher = {USENIX Association},
month = mar
}

@inproceedings{wei2023achelous,
  title={Achelous: Enabling programmability, elasticity, and reliability in hyperscale cloud networks},
  author={Wei, Chengkun and Li, Xing and Yang, Ye and Jiang, Xiaochong and Xu, Tianyu and Yang, Bowen and Wu, Taotao and Xu, Chao and Lv, Yilong and Gao, Haifeng and others},
  booktitle={Proceedings of the ACM SIGCOMM 2023 Conference},
  pages={769--782},
  year={2023}
}

@article{10.1145/2043164.2018465,
author = {Ballani, Hitesh and Costa, Paolo and Karagiannis, Thomas and Rowstron, Ant},
title = {Towards predictable datacenter networks},
year = {2011},
issue_date = {August 2011},
publisher = {Association for Computing Machinery},
address = {New York, NY, USA},
volume = {41},
number = {4},
issn = {0146-4833},
url = {https://doi.org/10.1145/2043164.2018465},
doi = {10.1145/2043164.2018465},
journal = {SIGCOMM Comput. Commun. Rev.},
month = aug,
pages = {242–253},
numpages = {12}
}

@inproceedings{10.1145/3689031.3717479,
author = {Wang, Xin and Li, Deguo and Wang, Zhihong and Jiang, Lidong and Wen, Shubo and Kang, Daxiang and Arslan, Engin and He, Peng and Qian, Xinyu and Niu, Bin and Pi, Jianwen and Ding, Xiaoning and Lin, Ke and Luo, Hao},
title = {Byte vSwitch: A High-Performance Virtual Switch for Cloud Networking},
year = {2025},
isbn = {9798400711961},
publisher = {Association for Computing Machinery},
address = {New York, NY, USA},
url = {https://doi.org/10.1145/3689031.3717479},
doi = {10.1145/3689031.3717479},
booktitle = {Proceedings of the Twentieth European Conference on Computer Systems},
pages = {1417–1432},
numpages = {16},
location = {Rotterdam, Netherlands},
series = {EuroSys '25}
}

@inproceedings{10.1145/3651890.3672224,
author = {Li, Xing and Jiang, Xiaochong and Yang, Ye and Chen, Lilong and Wang, Yi and Wang, Chao and Xu, Chao and Lv, Yilong and Yang, Bowen and Wu, Taotao and Gao, Haifeng and Chen, Zikang and Qiao, Yisong and Ding, Hongwei and Dong, Yijian and Yang, Hang and Song, Jianming and Lu, Jianyuan and Zhang, Pengyu and Wei, Chengkun and Zhang, Zihui and Chen, Wenzhi and He, Qinming and Zhu, Shunmin},
title = {Triton: A Flexible Hardware Offloading Architecture for Accelerating Apsara vSwitch in Alibaba Cloud},
year = {2024},
isbn = {9798400706141},
publisher = {Association for Computing Machinery},
address = {New York, NY, USA},
url = {https://doi.org/10.1145/3651890.3672224},
doi = {10.1145/3651890.3672224},
booktitle = {Proceedings of the ACM SIGCOMM 2024 Conference},
pages = {750–763},
numpages = {14},
location = {Sydney, NSW, Australia},
series = {ACM SIGCOMM '24}
}

@inproceedings{saeed2017carousel,
  title={Carousel: Scalable traffic shaping at end hosts},
  author={Saeed, Ahmed and Dukkipati, Nandita and Valancius, Vytautas and The Lam, Vinh and Contavalli, Carlo and Vahdat, Amin},
  booktitle={Proceedings of the Conference of the ACM Special Interest Group on Data Communication},
  pages={404--417},
  year={2017}
}

@inproceedings{MAPL,
  title={Network-aware placement of virtual machine ensembles using effective bandwidth estimation},
  author={Wang, Runxin and Esteves, Rafael and Shi, Lei and Wickboldt, Juliano Araujo and Jennings, Brendan and Granville, Lisandro Zambenedetti},
  booktitle={10th International Conference on Network and Service Management (CNSM) and Workshop},
  pages={100--108},
  year={2014},
  organization={IEEE}
}

@inproceedings{TIVC,
author = {Xie, Di and Ding, Ning and Hu, Y. Charlie and Kompella, Ramana},
title = {The only constant is change: incorporating time-varying network reservations in data centers},
year = {2012},
isbn = {9781450314190},
publisher = {Association for Computing Machinery},
address = {New York, NY, USA},
url = {https://doi.org/10.1145/2342356.2342397},
doi = {10.1145/2342356.2342397},
booktitle = {Proceedings of the ACM SIGCOMM 2012 Conference on Applications, Technologies, Architectures, and Protocols for Computer Communication},
pages = {199–210},
numpages = {12},
location = {Helsinki, Finland},
series = {SIGCOMM '12}
}

@inproceedings {295587,
author = {Biao Lyu and Enge Song and Tian Pan and Jianyuan Lu and Shize Zhang and Xiaoqing Sun and Lei Gao and Chenxiao Wang and Han Xiao and Yong Pan and Xiuheng Chen and Yandong Duan and Weisheng Wang and Jinpeng Long and Yanfeng Wang and Kunpeng Zhou and Zhigang Zong and Xing Li and Guangwang Li and Pengyu Zhang and Peng Cheng and Jiming Chen and Shunmin Zhu},
title = {{POSEIDON}: A Consolidated Virtual Network Controller that Manages Millions of Tenants via Config Tree},
booktitle = {21st USENIX Symposium on Networked Systems Design and Implementation (NSDI 24)},
year = {2024},
isbn = {978-1-939133-39-7},
address = {Santa Clara, CA},
pages = {1083--1099},
url = {https://www.usenix.org/conference/nsdi24/presentation/lyu},
publisher = {USENIX Association},
month = apr
}

@inproceedings{dalton2018andromeda,
  title={Andromeda: Performance, isolation, and velocity at scale in cloud network virtualization},
  author={Dalton, Michael and Schultz, David and Adriaens, Jacob and Arefin, Ahsan and Gupta, Anshuman and Fahs, Brian and Rubinstein, Dima and Zermeno, Enrique Cauich and Rubow, Erik and Docauer, James Alexander and others},
  booktitle={15th USENIX symposium on networked systems design and implementation (NSDI 18)},
  pages={373--387},
  year={2018}
}

@article{DBLP:journals/tsusc/JacquetLR24,
  author       = {Pierre Jacquet and
                  Thomas Ledoux and
                  Romain Rouvoy},
  title        = {{SCROOGEVM:} Boosting Cloud Resource Utilization With Dynamic Oversubscription},
  journal      = {{IEEE} Trans. Sustain. Comput.},
  volume       = {9},
  number       = {5},
  pages        = {754--765},
  year         = {2024}
}

@inproceedings{sc26disk,
  title={No Disk Left Behind: Lifetime-Aware Oversubscription of Thin Provisioned Storage in Cloud IaaS platforms},
  AUTHOR = {Jacquet, Pierre and Camille Coti and Marcos Dias de Assunção},
  booktitle={Proceedings of the International Conference for High Performance Computing, Networking, Storage and Analysis},
  year={2026}
}

@inproceedings {258850,
author = {Chunqiang Tang and Kenny Yu and Kaushik Veeraraghavan and Jonathan Kaldor and Scott Michelson and Thawan Kooburat and Aravind Anbudurai and Matthew Clark and Kabir Gogia and Long Cheng and Ben Christensen and Alex Gartrell and Maxim Khutornenko and Sachin Kulkarni and Marcin Pawlowski and Tuomas Pelkonen and Andre Rodrigues and Rounak Tibrewal and Vaishnavi Venkatesan and Peter Zhang},
title = {Twine: A Unified Cluster Management System for Shared Infrastructure},
booktitle = {14th USENIX Symposium on Operating Systems Design and Implementation (OSDI 20)},
year = {2020},
isbn = {978-1-939133-19-9},
pages = {787--803},
url = {https://www.usenix.org/conference/osdi20/presentation/tang},
publisher = {USENIX Association},
month = nov
}

@inproceedings{10.1145/3447786.3456259,
author = {Bashir, Noman and Deng, Nan and Rzadca, Krzysztof and Irwin, David and Kodak, Sree and Jnagal, Rohit},
title = {Take It to the Limit: Peak Prediction-Driven Resource over-commitment in Datacenters},
year = {2021},
isbn = {9781450383349},
publisher = {ACM},
OPTOPTaddress = {New York, NY, USA},
OPTurl = {https://doi.org/10.1145/3447786.3456259},
doi = {10.1145/3447786.3456259},
booktitle = {16th European Conference on Computer Systems},
pages = {556–573},
numpages = {18},
location = {Online Event, United Kingdom},
series = {EuroSys'21}
}

@inproceedings{verma2015large,
  title={Large-scale cluster management at Google with Borg},
  author={Verma, Abhishek and Pedrosa, Luis and Korupolu, Madhukar and Oppenheimer, David and Tune, Eric and Wilkes, John},
  booktitle={10th European Conference on Computer Systems},
series = {EuroSys'15},
  pages={1--17},
  year={2015}
}

@inproceedings{10.1145/3132747.3132772,
author = {Cortez, Eli and Bonde, Anand and Muzio, Alexandre and Russinovich, Mark and Fontoura, Marcus and Bianchini, Ricardo},
title = {Resource Central: Understanding and Predicting Workloads for Improved Resource Management in Large Cloud Platforms},
year = {2017},
isbn = {9781450350853},
publisher = {ACM},
OPTOPTaddress = {New York, NY, USA},
OPTurl = {https://doi.org/10.1145/3132747.3132772},
doi = {10.1145/3132747.3132772},
booktitle = {26th Symposium on Operating Systems Principles},
pages = {153–167},
numpages = {15},
location = {Shanghai, China},
series = {SOSP'17}
}

@article{jennings2015resource,
  title={Resource management in clouds: Survey and research challenges},
  author={Jennings, Brendan and Stadler, Rolf},
  journal={Journal of Network and Systems Management},
  volume={23},
  pages={567--619},
  year={2015},
  publisher={Springer}
}

@inproceedings{jacquet2023cloudfactory,
  title={CLOUDFACTORY: An Open Toolkit to Generate Production-like Workloads for Cloud Infrastructures},
  author={Jacquet, Pierre and Ledoux, Thomas and Rouvoy, Romain},
  booktitle={11th IEEE International Conference on Cloud Engineering},
series = {IC2E'23},
  year={2023}
}

@misc{sizingaws,
  author = {{AWS}},
  title = {AWS re:Invent 2017 Deep Dive on Amazon EC2 Instances, Featuring Performance Optimization (CMP301)},
  year = 2017,
  url = {https://www.youtube.com/watch?v=mZy6E2I5Rek&t=815s},
  note = {Available at \url{https://www.youtube.com/watch?v=mZy6E2I5Rek&t=815s}},
  OPTurldate = {2023}
}

@inproceedings {258979,
author = {Ori Hadary and Luke Marshall and Ishai Menache and Abhisek Pan and Esaias E Greeff and David Dion and Star Dorminey and Shailesh Joshi and Yang Chen and Mark Russinovich and Thomas Moscibroda},
title = {Protean: {VM} Allocation Service at Scale},
booktitle = {14th USENIX Symposium on Operating Systems Design and Implementation},
series = {OSDI'20},
year = {2020},
isbn = {978-1-939133-19-9},
pages = {845--861},
url = {https://www.usenix.org/conference/osdi20/presentation/hadary},
publisher = {USENIX Association},
month = nov
}

@misc{openstacksched,
  author = {{OpenStack}},
  title = {Scheduling},
  year = 2019,
  url = {https://docs.openstack.org/mitaka/config-reference/compute/scheduler.html},
  note = {Available at \url{https://docs.openstack.org/mitaka/config-reference/compute/scheduler.html}},
  OPTurldate = {2023}
}

@inproceedings{10.1145/3447786.3456225,
author = {Wang, Yawen and Arya, Kapil and Kogias, Marios and Vanga, Manohar and Bhandari, Aditya and Yadwadkar, Neeraja J. and Sen, Siddhartha and Elnikety, Sameh and Kozyrakis, Christos and Bianchini, Ricardo},
title = {SmartHarvest: Harvesting Idle CPUs Safely and Efficiently in the Cloud},
year = {2021},
isbn = {9781450383349},
publisher = {ACM},
OPTOPTaddress = {New York, NY, USA},
url = {https://doi.org/10.1145/3447786.3456225},
doi = {10.1145/3447786.3456225},
booktitle = {16th European Conference on Computer Systems},
pages = {1–16},
numpages = {16},
OPTlocation = {Online Event, United Kingdom},
series = {EuroSys'21}
}

@article{10.1145/2007477.1952709,
author = {Williams, Dan and Jamjoom, Hani and Liu, Yew-Huey and Weatherspoon, Hakim},
title = {Overdriver: handling memory overload in an oversubscribed cloud},
year = {2011},
issue_date = {July 2011},
publisher = {ACM},
OPTOPTaddress = {New York, NY, USA},
volume = {46},
number = {7},
issn = {0362-1340},
url = {https://doi.org/10.1145/2007477.1952709},
doi = {10.1145/2007477.1952709},
journal = {SIGPLAN Not.},
month = {mar},
pages = {205–216},
numpages = {12}
}

@inproceedings{fuerst2022memory,
  title={Memory-harvesting vms in cloud platforms},
  author={Fuerst, Alexander and Novakovi{\'c}, Stanko and Goiri, {\'I}{\~n}igo and Chaudhry, Gohar Irfan and Sharma, Prateek and Arya, Kapil and Broas, Kevin and Bak, Eugene and Iyigun, Mehmet and Bianchini, Ricardo},
  booktitle={Proceedings of the 27th ACM International Conference on Architectural Support for Programming Languages and Operating Systems},
  pages={583--594},
  year={2022}
}

@inproceedings{10.1145/3669940.3707226,
author = {Reidys, Benjamin and Zardoshti, Pantea and Goiri, \'{I}\~{n}igo and Irvene, Celine and Berger, Daniel S. and Ma, Haoran and Arya, Kapil and Cortez, Eli and Stark, Taylor and Bak, Eugene and Iyigun, Mehmet and Novakovic, Stanko and Hsu, Lisa and Trueba, Karel and Pan, Abhisek and Bansal, Chetan and Rajmohan, Saravan and Huang, Jian and Bianchini, Ricardo},
title = {Coach: Exploiting Temporal Patterns for All-Resource Oversubscription in Cloud Platforms},
year = {2025},
isbn = {9798400706981},
publisher = {Association for Computing Machinery},
address = {New York, NY, USA},
url = {https://doi.org/10.1145/3669940.3707226},
doi = {10.1145/3669940.3707226},
booktitle = {Proceedings of the 30th ACM International Conference on Architectural Support for Programming Languages and Operating Systems, Volume 1},
pages = {164–181},
numpages = {18},
location = {Rotterdam, Netherlands},
series = {ASPLOS '25}
}

@inproceedings {266915,
author = {Benjamin Hindman and Andy Konwinski and Matei Zaharia and Ali Ghodsi and Anthony D. Joseph and Randy Katz and Scott Shenker and Ion Stoica},
title = {Mesos: A Platform for {Fine-Grained} Resource Sharing in the Data Center},
booktitle = {8th USENIX Symposium on Networked Systems Design and Implementation (NSDI 11)},
year = {2011},
address = {Boston, MA},
url = {https://www.usenix.org/conference/nsdi11/mesos-platform-fine-grained-resource-sharing-data-center},
publisher = {USENIX Association},
month = mar
}

\end{document}